\documentclass[onecolumn,aps,prd,superscriptaddress,floats,floatfix,notitlepage]{revtex4}

\usepackage{graphicx,amssymb,amsmath,epsfig,mathtools,caption,csquotes} 
\usepackage{palatino}
\usepackage{xcolor}
\usepackage{hyperref}
\hypersetup{colorlinks=true,linkcolor=blue,urlcolor=blue,citecolor=blue}
\usepackage{tabularx}
\usepackage{booktabs}
\usepackage{multirow}
\usepackage{amsfonts}
\usepackage{array}
\usepackage{subcaption}
\usepackage{overpic}
\usepackage{titlesec}
\begin{document}
\title{Optical Properties and Thermal Geometries  of Hot NUT-Kerr-Newman-Kasuya-AdS Spacetime}
\author{M. Umair Shahzad}
\email{mushahzad@uo.edu.pk;m.u.shahzad@ucp.edu.pk}
\affiliation{Department of Mathematics, Faculty of Science, University of Okara, Okara 56130, Pakistan.}
\author{Nazek Alessa}
\affiliation{Department of Mathematical Sciences, College of Science,
Princess Nourah bint Abdulrahman University, P.O.Box 84428, Riyadh
11671, Saudi Arabia}

\author{Abdul Wahab}
\email{wahab.aryan@gmail.com}
\affiliation{Department of Mathematics, Faculty of Science, University of Okara, Okara 56130, Pakistan.}

\author{Rafda Rafique}
\email{rafdarafique86@gmail.com},
\affiliation{Department of Mathematics, Faculty of Science, University of Okara, Okara 56130, Pakistan.}
\affiliation{Western Caspian University, Baku, Azerbaijan.}

\date{\today}

\begin{abstract}
This paper is devoted to studying the optical and thermal geometrical properties of Hot,
NUT-Kerr-Newman-Kasuya-AdS black hole (BH). This BH is characterized by the NUT charge and a parameter $Q$ that comprises the electric and magnetic charge. We compute the image of the BH shadow in two types: 1) at \(\tilde{r}\to\infty \), 2) at \(\tilde{r}\to r_O \) by analytical approach. We also investigate the effect of Nut, spin, inclination angle, and cosmological constant on the shape of shadow. We analyze that for type 1, the shadow in increasing for higher values of NUT charge, the cosmological constant, rotation parameters, and inclination angle,    while for type 2, by increasing these parameters, the circular symmetry of the image of the BH shadow variate. Moreover, we discuss well-known thermal geometries such as Weinhold, Ruppeiner, HPEM, and  Quevedo case I \& II spacetime. It is found that Ruppeiner , HPEM and Quevedo (II) formulations provide physical information about the microscopic structure as compared to Weinhold and Quevedo (I) geometries of Hot NUT-Kerr-Newman-Kasuya-AdS BH. Our findings provides distinctive characteristics in the shadow and thermal geometries of this BH as compare to other BH types. 
\end{abstract}

\maketitle
\section{Introduction }
Einstein's General Relativity (GR) is a cornerstone of modern physics. This theory has successfully predict the phenomena that have been experimentally verified. Among its most intriguing predictions are black holes (BHs), gravitational monsters born from collapsed massive stars. In 1916, Karl Schwarzschild derived the first exact solution to Einstein’s field equations for a static, spherically symmetric vacuum now known as the Schwarzschild metric which describes the spacetime around a non-rotating BH \cite{p1}. Sir Arthur Eddington, during the famous total solar eclipse of 1919, provided crucial evidence supporting GR. His observations confirmed the bending of light by gravity, as predicted by Einstein’s theory \cite{p2}.\\
Cunningham and Bardeen \cite{p3} and Bardeen \cite{p4} have explored the optical signatures of BHs. They investigated scenarios such as a star orbiting a BH, shedding light on the fascinating interplay between gravity and light in these cosmic enigmas. They also proposed the basic geometry of a thin disk that has an inside zone which is gravitationally lensed. Luminet (1979) \cite{p5} showed handwritten images of a computerized picture of a BH encircled by a bright accretion disk. In 2000, Falcke et al. \cite{p6} coined the word ``shadow" and de Vries \cite{p7} after some weeks. In 2019, the Event Horizon Telescope (EHT)  collaboration had a remarkable success by presenting the primary image of a BH, captured from the supermassive BH M87* with the help of baseline interferometry technique \cite{r35,r36,r37,r38}. Since BH produces a visible shadow due to an intense gravitational field, the Schwarzschild BH is examined in the groundbreaking research of Luminet and Synge who also proposed a methodology to determine the angular radius of shadow \cite{p8,p5}.\\\
In 1963, rotating BHs, presented by the Kerr metric, yielded different shadow characteristics that were affected by their angular momentum. The spin parameter of rotating BH influenced the shadow's shape and dimensions. Later research by Bardeen  and Chandrasekhar \cite{p9} studied the properties of Kerr BH shadows, emphasizing the imbalance caused by rotation. Modern theoretical astrophysics has recently focused on AdS spinning BHs due to their profound implications for quantum and classical gravity. These BHs play a pivotal role in our understanding of the universe. Notably, BHs in AdS spacetimes exhibit unique features that are not found in asymptotically flat or dS spacetimes. For instance, AdS spacetimes are characterized by a negative cosmological constant, which profoundly influences their behavior. Also, BH shadow covers more unusual spacetimes, such as those with a NUT (Newman-Unti-Tamburino) charge. Grenzebach et al. (2014) \cite{p10} played a remarkable part in knowing how NUT charge effects the BHs. 
Moreover, the understanding of thermodynamic and geometrical features are critical in theoretical physics, especially for BHs and gravitational systems. The study of thermodynamic geometry models such as Ruppeiner, Weinhold, HPEM, and Quevedo reveal essential information on the fundamental structures of spacetime and their thermodynamic behaviors. Using these models, researchers may examine the correlations between entropy, temperature, and other thermodynamic variables, gain a better understanding of the fundamental features of BHs and other gravitational phenomena. Chabab et al. \cite{v1} provides the phase transition in charged AdS BHs by showing similarities in thermodynamic parameters sourced from Weinhold, Ruppeiner, and Quevedo metrics. This establishes a close connection between thermodynamics and statistics to focus on the importance of thermodynamic geometry in  thermodynamic systems \cite{v2}. The implementations  of geometrical methods in comprehending nonlinear BH solutions are demonstrated by applying thermodynamic geometry to Born-Infeld AdS BHs \cite{v3}. With a global monopole, researchers \cite{14A-53,v5} study phase transitions in charged AdS BHs using various thermodynamic geometries such as Weinhold, Ruppeiner, Quevedo, and HPEM formulations. \\
The aim of this paper is to investigate the shadow's empirical consequences and thermal  geometries of Hot-NUT-Kerr-Newman-Kasuya AdS BH (HNKNK-AdS BH) \cite{r2}, depending upon five important characteristics:  mass `M,' spin parameter `a=\(\frac{J}{M}\)', NUT `n', electric charge `\(Q_{e}\)',  magnetic charge `\(Q_{m}\)' and cosmological constant `\(\Lambda\)' in Boyer-Lindquist Coordinates (BLC) \cite{r3}. Furthermore, we look at the shadow of HNKNK-AdS BH might seem like to an observer at infinity and at a particular distance. Here, we assume there isn't a light source around the BH and we use geodesic light as the route of incident light rays. Moreover, we also investigate that the singularity of the curvature scalar in the Weinhold and Quevedo (case 1) formalism's doesn't match up with the divergence of the heat capacity which provides no physical information in that scenario. This difference suggests that the Weinhold and Quevedo (case 1) metrics might need to be physically relevant. On the other hand, the curvature scalars in the Ruppeiner, HPEM, and Quevedo (case 2) formalisms align with the zeros of heat capacity and divergence of curvature scalar. 

\section{The Null Geodesic Of Hot Nut-Kerr-Neman-Kasuya-Anti-De Sitter Black Hole }
A metric HNKNK-AdS BH \cite{r2} specifies with mass `M', spin parameter `a=\(\frac{J}{M}\)', NUT (magnetic mass) `n', the electric charge `\(Q_{e}\)', and magnetic mono-pole parameter  `\(Q_{m}\)' in BLC are given by \cite{r3}
\begin{equation}\label{1}
\begin{split}
ds^2 = & - \frac{(\Delta - a^2 \Delta_{\theta} \sin^2\theta)}{\Sigma} dt^2 
    + \frac{\Sigma}{\Delta} d\tilde{r}^2 
    + \frac{\Sigma}{\Delta_{\theta}} d\theta^2 \\
    & - \frac{(h^2 \Delta - \xi^2 \Delta_{\theta} \sin^2(\theta))}{\Xi^2 \Sigma} d\phi^2 
    - 2 \frac{(h \Delta - a \xi \Delta_{\theta} \sin^2(\theta))}{\Xi \Sigma} d\phi \, dt
 \end{split}
\end{equation}
where
\begin{equation}\label{2}
\begin{split}
    \Xi &= 1 - \frac{a^2}{y^2}, \\
    h &= a \sin^2\theta - 2 n \cos\theta, \\
    \Sigma &= \tilde{r}^2 + (n + a \cos\theta)^2, \\
    \Delta_{\theta} &= 1 - \frac{a^2 \cos^2(\theta)}{y^2}, \\
    \Delta &= \frac{\xi^2}{y^2} \left( 5 n^2 +\tilde{r}^2 + y^2 \right) - 2 M \tilde{r} - 2 n^2 + Q^2\\
    \xi&=\tilde{r}^2+a^2+n^2\\
    y^2&=-\frac{3}{\Lambda}
\end{split}
\end{equation}
where \(Q^2=Q^2_e +Q^2_m\).\\
 The Lagrangian formulation of the geodesic motion is used to obtain constant of motion connected to the spacetime symmetries
\begin{equation}\label{3}
    \mathcal{L} = \frac{1}{2} g_{\mu\nu} \frac{dx^\mu}{d\tau} \frac{dx^\nu}{d\tau}.
\end{equation}
The constant of motion E and L are given by using eqs. (\ref{1}) and (\ref{3})
\begin{equation}\label{4}
E = -\frac{\partial \mathcal{L}}{\partial \dot{t}} = -g_{t\phi} \dot{\phi} - g_{tt} \dot{t} = {\frac {  \Delta-{a}^{2}\Delta_{{\theta}} \left( {\it Sin}
 \left( \theta \right)  \right) ^{2} }{\Sigma}}\dot{t}+{\frac {h \Delta-a\xi\,\Delta_{{\theta}} \left( {\it Sin} \left( \theta \right)  \right) ^{2}} {\Xi\,\Sigma}}\dot{\phi},
 \end{equation}
 \begin{equation}\label{4a}
L_z = \frac{\partial \mathcal{L}}{\partial \dot{\phi}} = g_{\phi\phi} \dot{\phi} + g_{t\phi} \dot{t}=-\frac{h^2 \Delta - \xi^2 \Delta_{\theta} \left( {\it Sin} \left( \theta \right)  \right) ^{2}}{\Xi^2 \Sigma} \dot{\phi} - \frac{h \Delta - a \xi \Delta_{\theta} \left( {\it Sin} \left( \theta \right)  \right) ^{2}}{\Xi \Sigma} \dot{t}.
\end{equation}
These two conserved values eqs. (\ref{4}) and (\ref{4a}) allow us to write the null geodesic equations as \cite{r4}
\begin{equation}\label{5}
   \dot{\phi}=-{\frac {{\Xi}^{2}L{a}^{2}+\Xi\,Ea\xi}{\Delta\,\Sigma}}+{\frac {\Xi\,E
h+{\Xi}^{2}L}{ \left( {\it Sin} \left( \theta \right)  \right) ^{2}
\Delta_{{\theta}}\Sigma}},
\end{equation}
\begin{equation}\label{5a}
   \dot{t}=-{\frac {L\Xi\,a\xi+E{\xi}^{2}}{\Delta\,\Sigma}}+{\frac {E{h}^{2}+L\Xi
\,h}{\Sigma\, \left( {\it Sin} \left( \theta \right)  \right) ^{2}
\Delta_{{\theta}}}}.
\end{equation}
The Hamilton-Jacobi equation is used to get the remaining geodesic equations for the rotating BH with metric tensor:
\begin{equation}\label{6}
\frac{\partial S}{\partial \sigma} = -\frac{1}{2} g^{\mu\nu} \frac{\partial S}{\partial x^\mu} \frac{\partial S}{\partial x^\nu}.
\end{equation}
The solution to the Hamilton-Jacobi equation:
\begin{equation}\label{7}
S = \frac{1}{2} m_0^2 \sigma - E t + L_z \phi + S_{\tilde{r}} + S_\theta,
\end{equation}
\( m_{0} \) represents particle's mass at rest (zero in the case of null geodesics). One can find the angular and radial equations 
\begin{equation}\label{8}
   \Sigma^2 \dot{\theta}^2 =\Theta(\theta) = \Delta_{\Theta} K - \frac{1}{{\sin^2 \theta}}(hE + L\Xi)^2,
   \end{equation}
   \begin{equation}\label{8a}
   \Sigma^2 \dot{\tilde{r}}^2 =
   R(\tilde{r}) = (\xi E + a L \Xi)^2 - \Delta_r K,
\end{equation}
where K is the \textit{Carter constant} \cite{r1}, \(R(\tilde{r})\) and \(\Theta (\theta)\) are the angular and radial coordinate's functions respectively.
\begin{equation}\label{9}
    \zeta=\frac{L}{E}\\,\eta=\frac{K}{E^2},
\end{equation}
by providing these conditions, we can find BH shadow
 \begin{equation}\label{10}
    R(\tilde{r}) = 0 \quad \text{and} \quad R \,'(\tilde{r}) = 0.
 \end{equation}
 Now using eq. (\ref{8a}), (\ref{9}) and (\ref{10}), dimensionless parameters are as follows:
 \begin{equation}\label{11}
      \zeta =-\frac{{3(3a^2 M + 3M n^2 + 3a^2 \tilde{r} - 9n^2 \tilde{r} + 6Q^2 \tilde{r} - 9M\tilde{r}^2 + 3\tilde{r}^3 + N\Lambda)}}{{a(3 + a^2 \Lambda)(3M - 3\tilde{r} + a^2 \tilde{r} \Lambda + 6n^2 \tilde{r} \Lambda + 2\tilde{r}^3 \Lambda)}},
      \end{equation}
      \begin{equation}\label{11a}
      \begin{split}
      \eta &=-\frac{{12 \tilde{r}^2 (-3a^2 + 3n^2 - 3Q^2 + 6M\tilde{r} - 3\tilde{r}^2 + 5a^2 n^2 \Lambda + 5n^4 \Lambda + a^2\tilde{r}^2 \Lambda + 6n^2 \tilde{r}^2 \Lambda + \tilde{r}^4 \Lambda)}}{{(3M - 3\tilde{r} + a^2 \tilde{r} \Lambda + 6n^2 \tilde{r} \Lambda + 2\tilde{r}^3 \Lambda)^2}},\\
      where\\
      N=&a^4 \tilde{r}- 3a^2 n^2 \tilde{r}- 4n^4 \tilde{r}+ a^2 \tilde{r}^3- 4n^2 \tilde{r}^3.
      \end{split}
 \end{equation}
  \section{Shadow Of Black Hole For an Observer at 
 \texorpdfstring{$\tilde{r} \to \infty $}{\tilde{r} to infinity}}
\begin{figure}
    \centering
    \begin{tabular}{cccc}
\includegraphics[width=5cm]{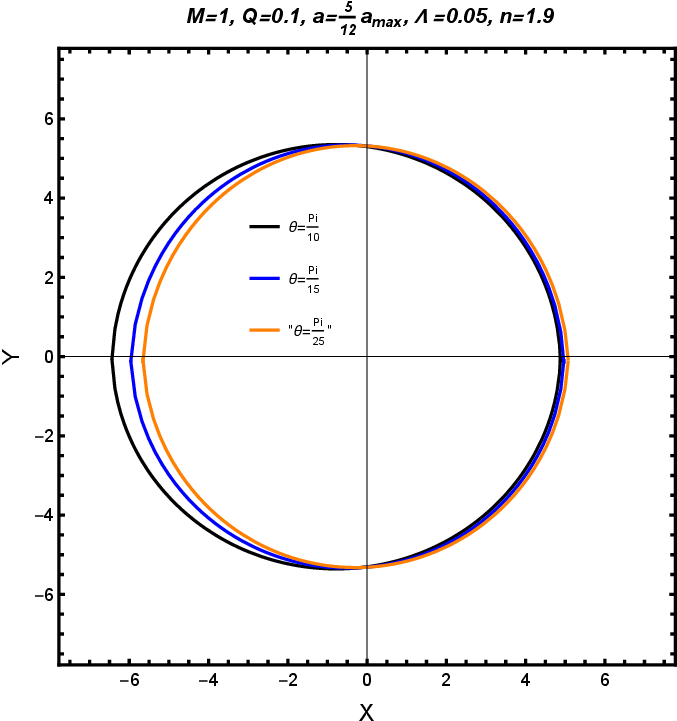}& 
\includegraphics[width=5cm]{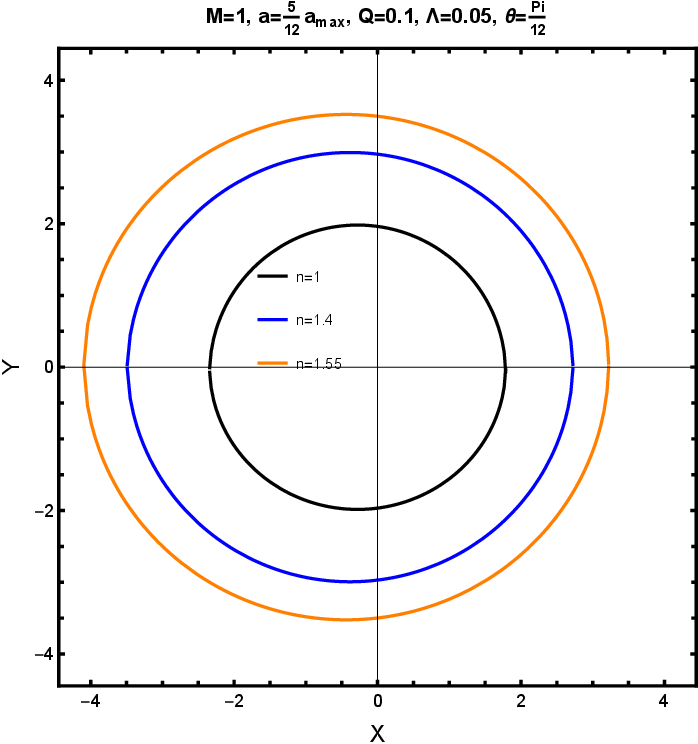}&
\includegraphics[width=5cm]{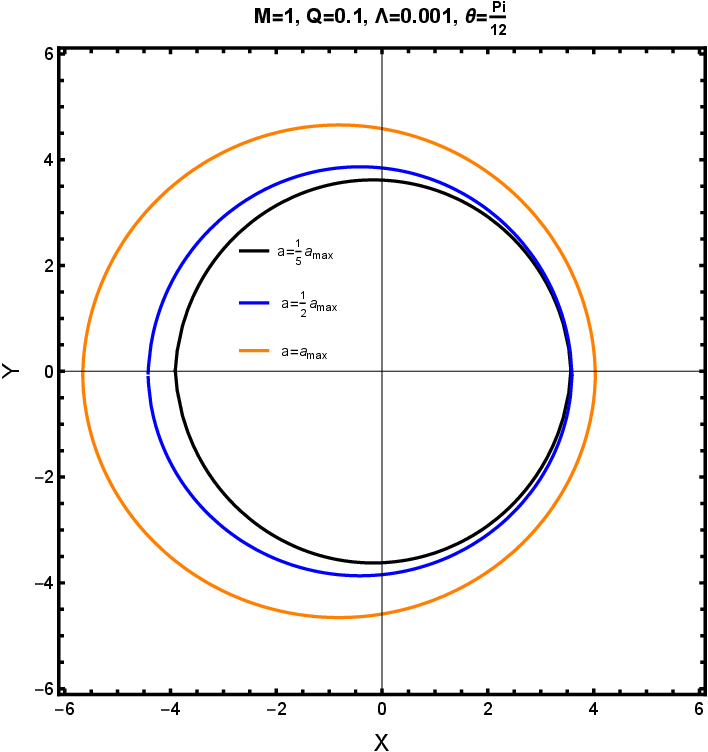}\\
\textbf{(a)}  & \textbf{(b)} & \textbf{(c)}\\[6pt]
\end{tabular}
\begin{tabular}{cccc}
\includegraphics[width=5cm]{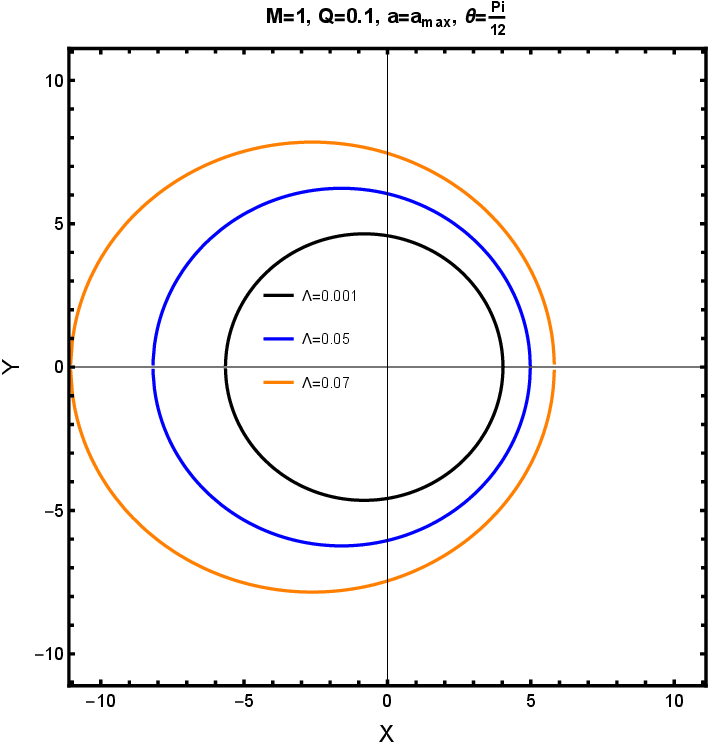} \\
\textbf{(d)} \\[6pt]
\end{tabular}
\caption{ \textbf{(a)} BH shadow decreases as values of inclination angle decrease, here \(a_{\text{max}}\)=2.145
\textbf{(b)} BH shadow becomes larger with increase values of NUT
\textbf{(c)} BH shadow smaller in size with spin variation a=\(\frac{1}{5}\)\(a_{\text{max}}\)=0.43,\(\frac{1}{3}\)\(a_{\text{max}}\)=0.73,\(\frac{1}{2}\)\(a_{\text{max}}\)=1.07,\(a_{\text{max}}\)=2.1 
\textbf{(d)} shadow of given BH increases by increasing values of cosmological constant}
\label{fig:1}
\end{figure}

   This section investigates the HNKNK-AdS BH's shadow observing at \( \tilde{r}\to\infty \).  In literature, we see that on the equatorial plane of the BH, a photon cannot maintain a circular trajectory with a constant radius and latitude angle because of the existence of the NUT charge \cite{r5,r6,r7}. As such, it produces a shadow profile different from that of a BH, in which a photon may continue to orbit in an equatorial plane. The mathematical description of the celestial coordinates \cite{r10} is
\begin{equation}\label{12}
\alpha = \lim_{{\tilde{r}_0 \to \infty}} \left( \int_{-\tilde{r}_0}^{0} \frac{-\tilde{r}}{2} \sin \theta_0 \frac{d\phi}{d\tilde{r}} \, d\tilde{r} \right)
\end{equation}
\begin{equation}\label{13}
\beta = \lim_{{\tilde{r}_0 \to \infty}} \left( \int_{\tilde{r}_0}^{0} \frac{d\theta}{d\tilde{r}} \, d\tilde{r} \right)\
\end{equation}
where $\theta_{0}$ is the latitude angle between the symmetrical axis and the normal observer, while $\tilde{r}_{0}$ is the separation between the observer and the massive object. Using the geometric formalism of the straight line linking the apparent location of the picture to the position of the distant observer in order to  write ($\alpha, \beta$) in the Euclidean coordinates system, transfer it to spherical coordinates and deduce eqs. (\ref{12}) and (\ref{13}) as
\begin{equation}\label{14}
\begin{split}
\alpha &=\frac{\zeta}{Sin(\theta_0)},\\
\beta &=\sqrt{\eta^2+a^2{(cos^2{(\theta_0)}}-\zeta^2{cot^2{(\theta_0)}}}
\end{split}
\end{equation}
If $\Lambda \leq 0$ then $\Delta$ has positive second order derivatives respective to $\tilde{r}$ and roots of  $\Delta$ are either 2 or 0 in which BH is in the first case while regular spacetime is in the second case. In case of BH, domain of outer communication is the area between $\tilde{r} \to \infty$ and first horizon and one can consider the observer in this area for viewing the BH shadow. $\partial_r$ behaves space-like on the domain of outer communication, which is analogous to $\Delta > 0$.
At \(\Lambda = 0\),  \(\Delta\) reduces to a second-order equation in terms of \(\tilde{r}\) and for an observer at infinity which is
\begin{equation}\label{15}
\begin{split}
    \Delta &= \frac{(\tilde{r}^2+a^2+n^2)^2}{y^2} \left( 5 n^2 +\tilde{r}^2 + y^2 \right) - 2 M \tilde{r} - 2 n^2 + Q^2,\\
    y^2 &=\frac{-3}{\Lambda}.
    \end{split}
\end{equation}
Thus, by using eq. (\ref{15}), one can obtain the horizon radius 
\begin{equation}\label{16}
\tilde{r}_{\pm}=M\pm\sqrt {{M}^{2}-{Q}^{2}-{a}^{2}+{n}^{2}}
\end{equation}
If \(a^2 \leq a^2_{\text{max}}\):= ${M}^{2}-{Q}^{2}+{n}^{2} $; If \(a^2 > a^2_{\text{max}}\), we have no horizons (regular space time rather than a BH). When the cosmological constant $\Lambda$ is not zero, spacetime does not remain asymptotically flat. Consequently, an observer within the domain of outer communication is separated from an observer at infinity by a cosmological horizon when $\Lambda$ is greater than zero \cite{r9,r10,r44} \\
Taking into account the above mentioned circumstances, we then plot \(\beta\) as a function of \(\alpha\) by using eq. (\ref{14}) to get the BH shadow. In fig. \ref{fig:1} these charts are shown for various parameters such as spin `a', Nut charge `n', inclination angle `$\theta$' and cosmological constant `$\Lambda$'.
It is evident that, as values of the Nut charge and spin parameter increase, the shadow increases while for decreasing inclination angle the shadow becomes distorted. Additionally, an increase in the cosmological constant leads to a larger shadow.
\section{Shadow Of Black Hole For An Observer At \texorpdfstring{$\tilde{r} = r_{O}$}{\tilde{r} to rO}}
This section investigate the BH shadow from the perspective of \( \tilde{r} = r_O \). We analyze spherical lightlike geodesics that remain on a sphere of constant radius, known as the `photon region'. Photon region actually is fundamentally construction of an image of shadow. Therefore, Our observer is located at coordinates (\( {r}_{O},\theta_{O} \)) in domain of outer communication, where \( \Lambda > 0 \) and by considering a light source at \( \tilde{r} = r_L \) with \( r_L \geq r_O \). Now we calculate the BH shadow through these circumstances: Firstly, we determine the ``Orthonormal-tetrad" at our observer's domain \cite{r10}
\begin{figure}
\centering
\includegraphics[width=8cm]{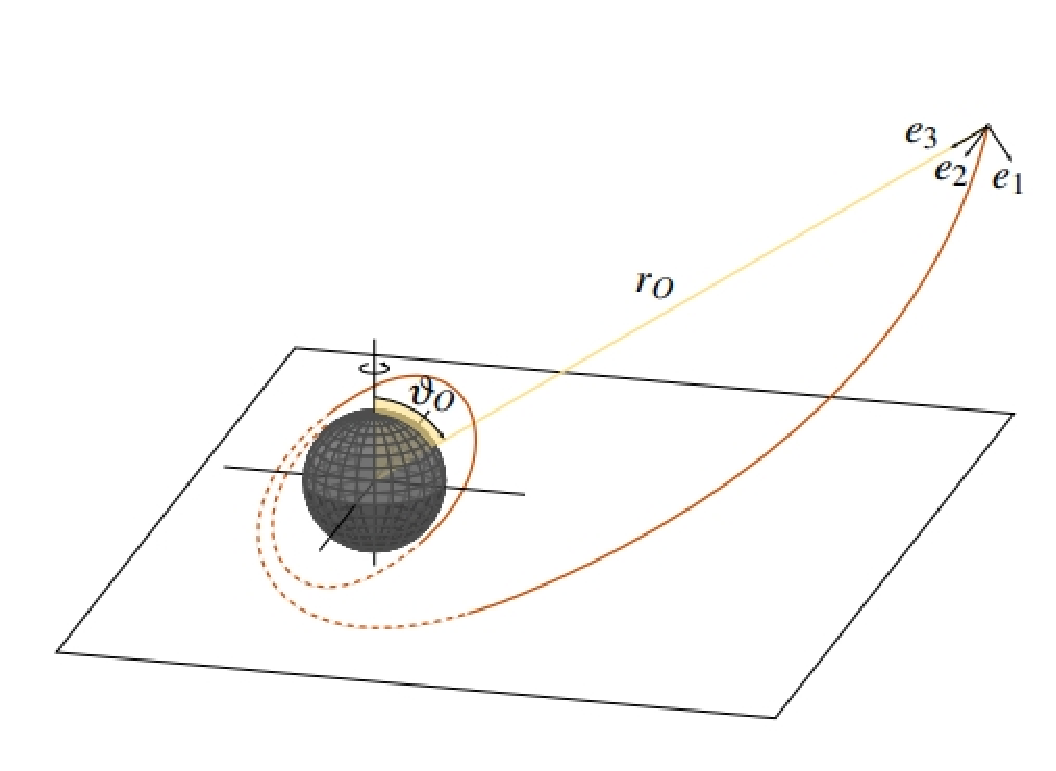} 
\caption{ At the observation event with BLC coordinates (\(r_O, \theta_O\)) we choose an orthonormal-tetrad following the prescriptions in eqs. (\ref{17}). A linear combination of these basis vectors, namely (\(e_{0}, e_{1}, e_{2}, e_{3}\)), can be employed to describe the tangent vector for each light beam transmitted from the observational event backward in time \cite{r10}.}\label{Fig.2}
\end{figure}
\begin{equation}\label{17}
\begin{split}
    (e_0) & : \quad e_0 = \frac{\frac{(a^2 + \tilde{r}^2+n^2)}{\Xi} \partial_t + a\partial_{\varphi}}{\sqrt{\Sigma} \sqrt{\Delta}} \bigg|_{(r_O, \theta_O)} \\
    (e_1) & : \quad e_1 = \frac{\sqrt{\Delta_\theta}}{\sqrt{\Sigma}} \partial_\theta \bigg|_{(r_O, \theta_O)} \\
    (e_2) & : \quad e_2 = -\left( \frac{\frac{h}{\Xi} \, \partial_t + \partial_{\varphi}}{\sqrt{\Delta_\theta} \sqrt{\Sigma}\sin \theta} \right) \bigg|_{(r_O, \theta_O)} \\
    (e_3) & : \quad e_3 = -\frac{\sqrt{\Delta}}{\sqrt{\Sigma}} \partial_{\tilde{r}} \bigg|_{(r_O, \theta_O)}
\end{split}
\end{equation}
Since we take the area of space where communication is feasible for an observer so that both \(\Sigma\) and \(\Delta\) are positive in this area. Moreover, we restrict our values of `a' and `$\Lambda$' such that $\Delta_\theta$ must be positive.
As a result, eq. (\ref{17}) have real coefficients and it is simple to confirm that (\(e_{0}, e_{1}, e_{2}, e_{3})\) are orthonormal. The spatial orientation of the BH's center is shown by the vector $e_3$ (see fig: \ref{Fig.2} \cite{r10}). The four-velocity of our observer is represented by the time-like vector \(e_{0}\). The tetrad is selected so that the primary null directions of metric (\ref{1}) are aligned with \(e_{0} \pm e_{3}\). An observer with four-velocity \(e_{0}\) is point toward the center of BH with vector \(e_{3}\).
For every light beam  \(\Upsilon(s)\), we have coordinates are \(\tilde{r}(s)\), \(\theta(s)\), \(\phi(s)\) and t(s). As a result, the following is an expression for the tangential vector where an observer is situated:
\begin{equation}\label{18}
   \dot \Upsilon = \tilde{r}\,\partial_{\tilde{r}} + \dot{\theta}\,\partial_\theta + \dot{\phi}\,\partial_\phi + \dot{t}\,\partial_t\
\end{equation}
\begin{equation}\label{19}
 \dot \Upsilon  = \Omega\, \left( -\,e_0 + sin \omega \,cos\psi \,e_1 + sin \omega \,sin \psi\, e_2 + cos \theta \,e_3 \right)
\end{equation}
where the scalar factor \(\Omega\) can be found by using eqs. (\ref{1}) and (\ref{4})
\begin{equation}\label{20}
   \Omega= -{\frac {L\Xi\,a+E\xi}{\Xi\,\sqrt {\Delta}\sqrt {\Sigma}}}
\end{equation}
Here ``celestial-coordinates " \(\omega\) and \(\psi\) for an observer are given by equation (\ref{19}) as shown in fig.\ref{Fig.3} and
\begin {figure}
\centering
\includegraphics[width=10cm]{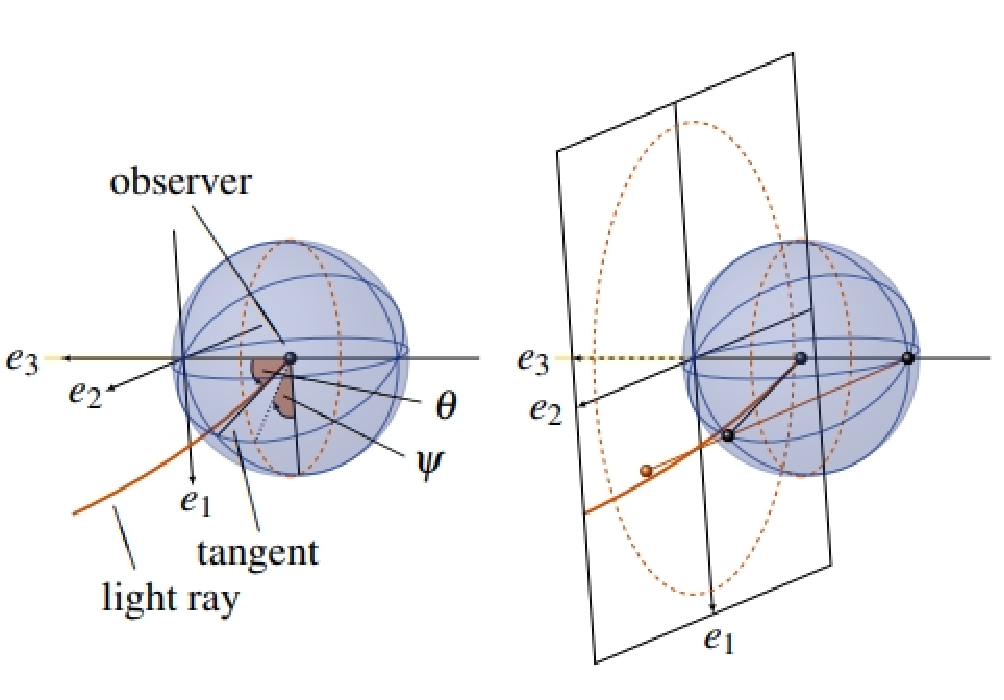} 
\caption{ For every light beam, we chose  \(\omega\) and \(\psi\) from equation (\ref{19}). The stereographic projection of points ($\theta,\psi$) onto the celestial plane is depicted by the red (color) dot in the right image. Additionally, the celestial equator's projection \(\theta=\frac{\pi}{2}\) is illustrated by the red dotted circles in the same image \cite{r10} .}\label{Fig.3}
\end{figure}
by comparing coefficients of \(\partial_{\phi}\) and \(\partial_{\tilde{r}}\) from eqs. (\ref{18}) and (\ref{19}) we get,
\begin{equation}\label{21}
\sin \psi = -\frac{\sqrt{\Delta_{\theta}}  \sqrt{\Sigma}\,\sin\theta}{\sin\omega} \left( \frac{\dot\varphi}{\Omega} + \frac{a}{\sqrt{\Delta}  \sqrt{\Sigma}} \right) \bigg|_{(r_O, \theta_O)} 
\end{equation}
\begin{equation}\label{22}
    \cos\omega = -\frac{\sqrt{\Sigma}\, \dot {\tilde{r}}}{\sqrt{\Delta} \Omega}\bigg|_{(r_O, \theta_O)}
\end{equation}
Using eqs. (\ref{5}),(\ref{8a}) and (\ref{9}), we obtain eqs. (\ref{23}) and (\ref{24}) as follows:
\begin{equation}\label{23}
   \sin\psi = -{\frac { \left( \Delta_{{\theta}} \left( {\Xi}^{3}a+h \right)  \left( 
\zeta \Xi\,a+\xi \right)   sin^{2}  \theta    
-{\Xi}^{3}\Delta\, \left( h+\zeta\Xi \right)  \right) \sqrt {\Delta_{
{\theta}}\Sigma}}{\sqrt {\Delta\,\Sigma} sin  \theta  sin  \omega \Delta_{{\theta}} \left( \zeta\Xi
\,a+\xi \right) \Xi}}
\bigg|_{(r_O, \theta_O)} 
\end{equation}
\begin{equation}\label{24}
    \cos\omega = \frac{\sqrt{( \Xi \zeta a + \xi)^2 - \Delta \eta} \,\Xi}{\Xi \zeta a + \xi}\bigg|_{(r_O, \theta_O)}
\end{equation}

\begin{equation}\label{25}
sin\omega=\sqrt {1-{\frac { \left((  \Xi\, \zeta \,a+\xi\right) ^{2}-\Delta\,\eta ) {\Xi}^{2}}{ ( \Xi\, 
\zeta  \,a+\xi)  ^{2}}}}\bigg|_{(r_O, \theta_O)}
\end{equation}
 Next, we plot the images of the BH shadow by using eqs. (\ref{11}), (\ref{11a}), (\ref{23}) to (\ref{25}) and the stereographic projection from celestial plane (see fig: \ref{Fig.3} \cite{r10}). Eqs. (\ref{23}) and (\ref{25}) in particular, specify the shape of the shadow the BH. Indeed, the shadow's counter showed the light rays that approach the spherical light like geodesic of radius \( \tilde{r}_p \) and motion parameters of these light beams must have the same as limiting spherical light like geodesic. Furthermore, the following methods yield the cartesian coordinates in terms of celestial coordinates:
 \begin{equation}\label{27}
     {X}(\tilde{r}_p) = 2 \tan\left(\frac{\omega(\tilde{r}_p)}{2}\right) \sin\left(\psi(\tilde{r}_p)\right)
 \end{equation}
 \begin{equation}\label{28}
    {Y}(\tilde{r}_p) = 2 \tan\left(\frac{\omega(\tilde{r}_p)}{2}\right) \cos\left(\psi(\tilde{r}_p)\right)
 \end{equation}
 \begin{figure}
    \centering
    \begin{tabular}{cccc}
\includegraphics[width=5cm]{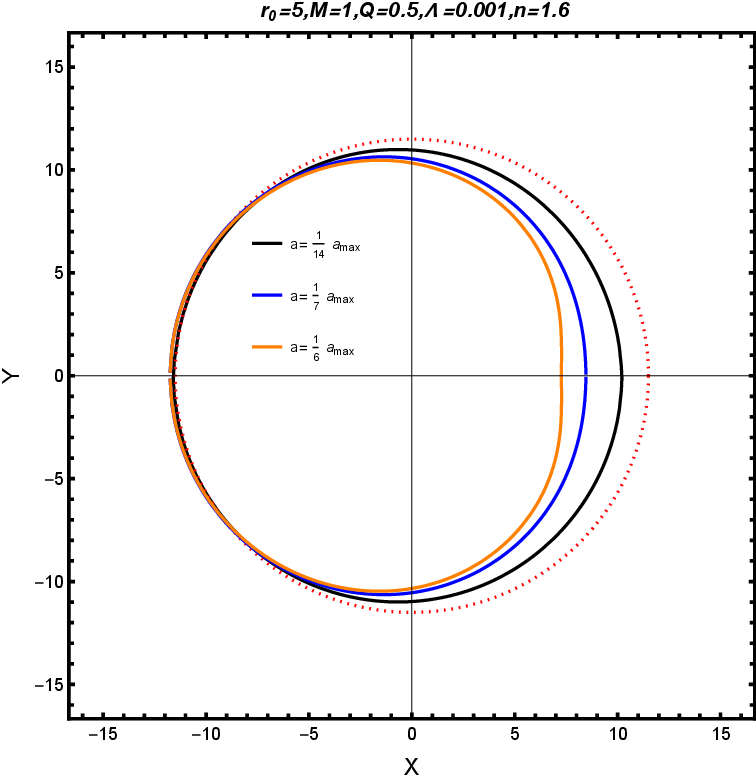}& 
\includegraphics[width=5cm]{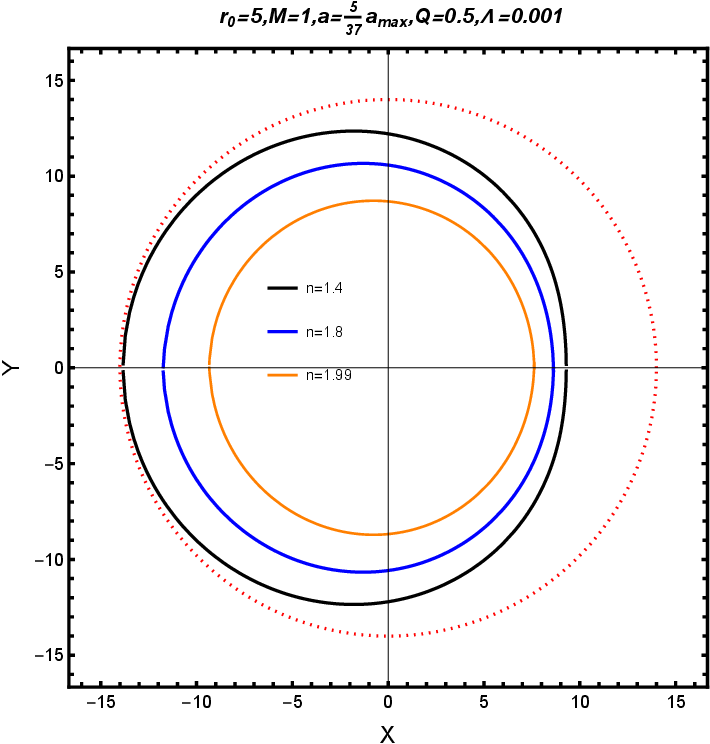}&
\includegraphics[width=5cm]{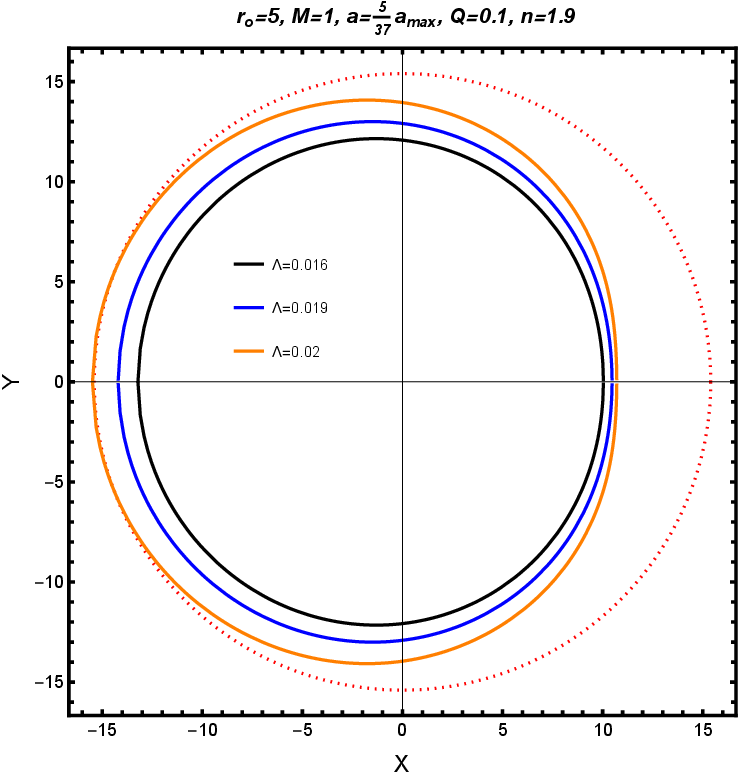}\\
\textbf{(a)}  & \textbf{(b)} & \textbf{(c)}\\[6pt]
\end{tabular}
\caption{ Plot illustrating shadow size of HNKNK-AdS BH with various spin,nut and cosmological constant values and other fixed parameters, for (a), (b) and (c) respectively.The red dashed circle shows the celestial equator (as a reference circle)}
\label{fig:4}
\end{figure}
The photon area or shadow limit of the revolving BH see fig.\ref{fig:4} is observed by a stagnant observer at (\( \tilde{r}= r_{O}, \theta_{O} = \frac{\pi}{2}\)) in the plot of  X vs.Y from eqs. (\ref{27}) and (\ref{28}). The BH characteristics are chosen carefully to keep the observer inside the outer communication zone. The shadow is shown for four distinct spin values; \(a = \lambda\, a_{\text{max}}\), where \( \lambda = \frac{1}{14}, \frac{1}{7}, \frac{1}{6}\) and
 \(a_{\text{max}}\) is defined by eq. (\ref{16}). A shadow changes to D-shaped as `\(a\)' rises. 
 As the Nut parameter ‘n’ increases, the shadow size decreases. Additionally, an increase in the cosmological constant results in a larger shadow shape, while keeping other parameters fixed 
 see figure \ref{fig:4} part (b),(c) and (d), respectively. 
\section{Thermal  Geometries Of HNKNK-AdS BH} 
In this section, first, we examine thermodynamic properties , such as `mass', `temperature,' and `heat capacity'. 
The temperature, entropy, and specific heat capacity indicate phase transitions and critical points, demonstrating variation in the BH nature and stability under different conditions. The event horizon is determined using eq. (\ref{1}), and the mass function is derived by considering the condition  $ \Delta = 0$. 
Following this, the corresponding values for temperature, entropy , and specific heat capacity are computed \cite{r2}

\begin{eqnarray}\label{29}
T = \frac{\Delta' \Xi}{4\pi(a^2 + n^2 + \tilde{r}_+^2) }
 \end{eqnarray}
\begin{eqnarray}\label{30}
S = \frac{\pi(a^2 + n^2 + \tilde{r}_+^2) }{\Xi }
 \end{eqnarray}
The specific heat capacity is given by
 \begin{eqnarray}\label{31}
C_{H} = (\frac{\partial M }{\partial \tilde{r}_+ })(\frac{\partial \tilde{r}_+}{\partial T })
 \end{eqnarray}
where $ \tilde{r}_+ $ showing the event horizon 
\begin{figure}
\centering
\includegraphics[width=0.50\textwidth]{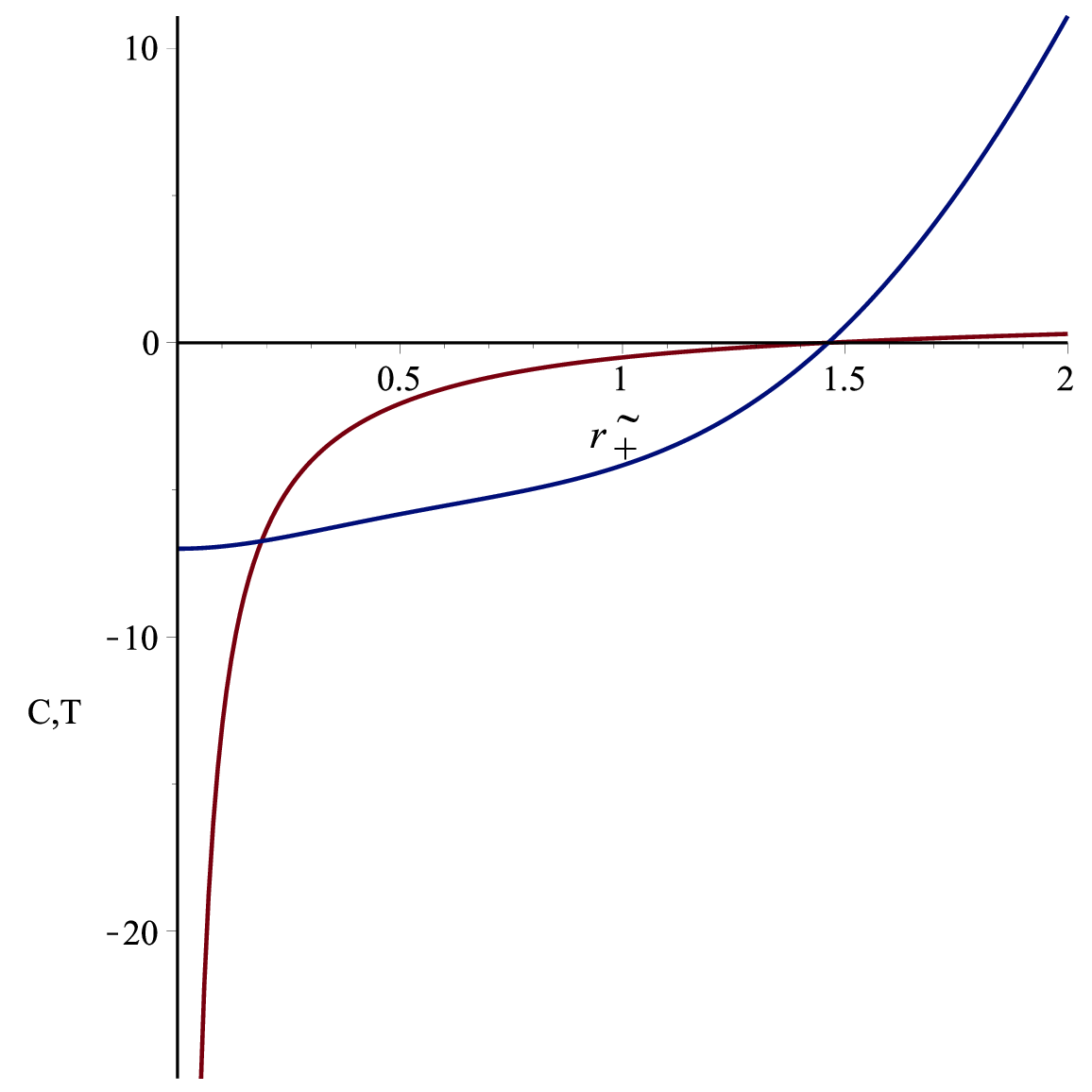}
\caption{Heat capacity and temperature  at Q = 4.3, n = 0.79, a = 0.7, y = 97}\label{Fig:5}
\end{figure}
In fig. \ref{Fig:5}, the heat capacity and temperature is plotted against the radial coordinates. The zeros of heat capacity and temperature coincide at $ \tilde{r}_+ = 1.4 $ which provides useful information. Furthermore, at  $ \tilde{r}_+ < 1.4 $, the system is unstable as well as nonphysical and at  $ \tilde{r}_+ > 1.4 $, the system becomes stable and physical. Hence, one can conclude that $ \tilde{r}_+ = 1.4 $ is a phase transition point.

Moreover, we explore thermal geometries by utilizing Weinhold, Ruppiner, Quevedo's (case I and II) and  HPEM spacetime.In recent years, Ruppeiner, Weinhold, Quevedo, and HPEM metrics have been employed to thoroughly quantify rotating black holes (BHs) \cite{14A-53,9A-60,10A-1-12,13A- 130,7A- 05,15A- 21}. 
 \begin{equation}\label{32}
 g^W_{ij} = \partial_{i}\partial_{j}M(S,Q,a)
 \end{equation}
 The following is the expression for the Weinhold geometry and its line element.
\begin{equation}
\begin{aligned}\label{33}
ds^2_W &= \left( \frac{\partial^2 M}{\partial S^2} \right) dS^2 
        + \left( \frac{\partial^2 M}{\partial Q^2} \right) dQ^2 
        + \left( \frac{\partial^2 M}{\partial a^2} \right) da^2 \\
        &\quad + 2 \left( \frac{\partial^2 M}{\partial S \partial Q} \right) dS dQ 
        + 2 \left( \frac{\partial^2 M}{\partial a \partial Q} \right) da dQ 
        + 2 \left( \frac{\partial^2 M}{\partial S \partial a} \right) dS da
\end{aligned}
\end{equation}

 \begin{eqnarray} \label{34}
 d{s_R}^2 = -\frac{1}{T} d{s_W}^2
  \end{eqnarray}
Ruppeiner geometry is a form of information geometry used to study thermodynamics. It suggests that thermodynamic systems can be represented using Riemannian geometry and statistical properties can be derived from this model. This geometrical model is based on the idea that there exist equilibrium states which can be represented by points on  surface and the distance between these equilibrium states is related to the fluctuation between them. The line element in summarize way is given by
 \begin{eqnarray}\label{35}
ds^2&=&~\begin{cases}
Mg^W_{ab}dX^b dX^a \quad~~~~~~~~~~~~~~~~~~~~~~~~~~~~~~~~~~~~~~          \text{Weinhold} &\ \ \\
-T^{-1}Mg^W_{ab}dX^b dX^a \quad~~~~~~~~~~~~~~~~~~~~~~~~~~~~~~~     \text{Ruppeiner} &\ \ \\
\frac{SM_S}{(M_{QQ}M_{aa})^3}(-M_{SS}dS^2 + M_{QQ}dQ^2 + M_{aa}da^2)\quad \text{HPEM}  &\ \ \\
{SM_S}(-M_{SS}dS^2 + M_{QQ}dQ^2 + M_{aa}da^2) ~~~~~~\quad \text{Quevedo case I} &\ \ \\
({SM_S}+{QM_Q}+{aM_a})(-M_{SS}dS^2 + M_{QQ}dQ^2 \ \ \\+ M_{aa}da^2)~~~~~~~~~~~~~~~~~~~~~~~~~~~~~~~~~~~~~\quad \text{Quevedo case II} &\ \ \\
\end{cases}
\end{eqnarray}
\begin{figure}
\centering
\begin{minipage}[h]{0.43\textwidth}
\includegraphics[width=\textwidth]{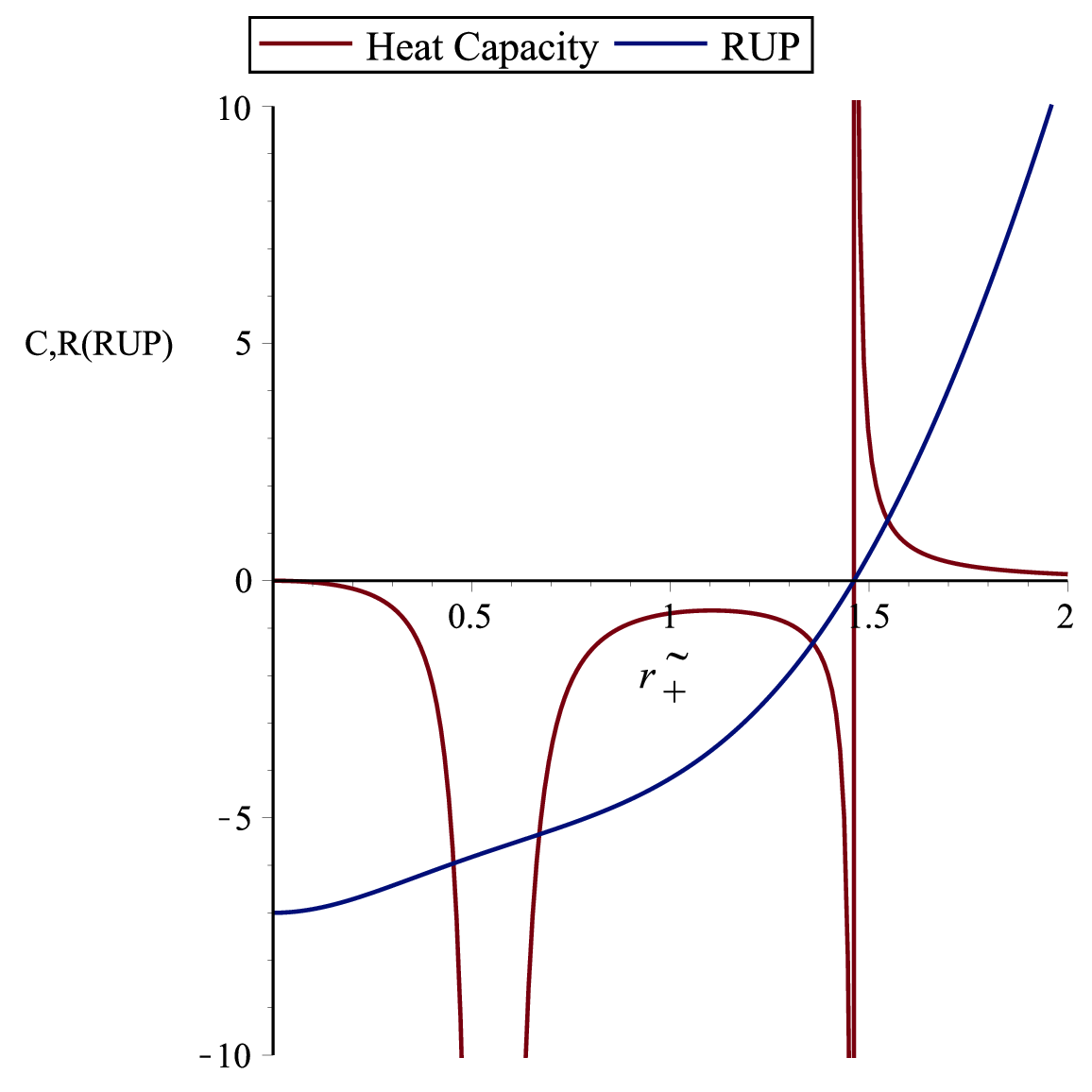}
\caption{Heat capacity and RUP Metric at Q = 4.3,n =0.79 ,a = 0.7 ,y = 97}\label{Fig:6}
\end{minipage}
\hfill
\begin{minipage}[h]{0.43\textwidth}
\includegraphics[width=\textwidth]{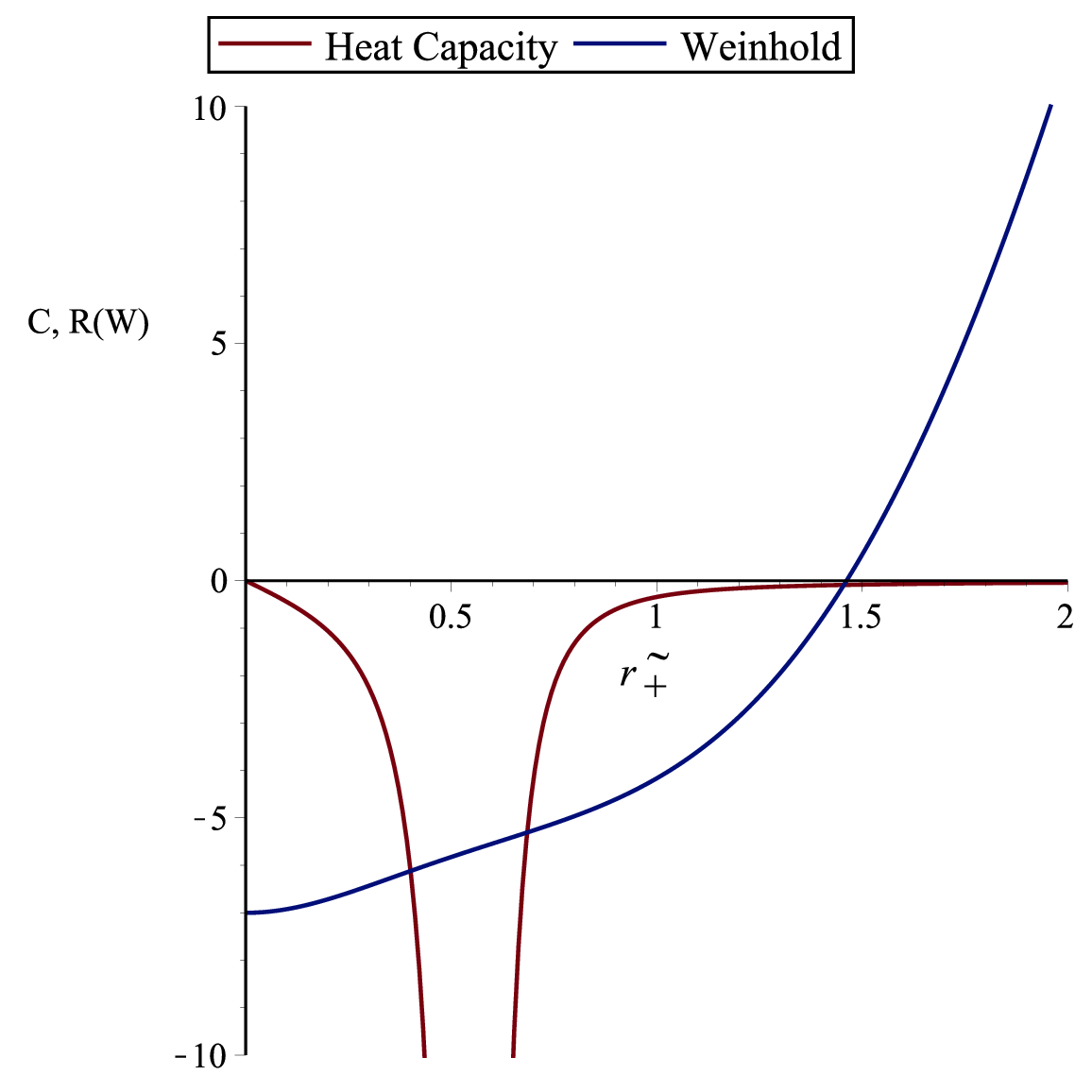}
 \caption{Heat capacity and Weinhold Metric at  Q = 4.3,n =0.79 ,a = 0.7 ,y = 97}\label{Fig:7}
\end{minipage}
\end{figure}

\begin{figure}
\centering
\begin{minipage}[h]{0.43\textwidth}
\includegraphics[width=\textwidth]{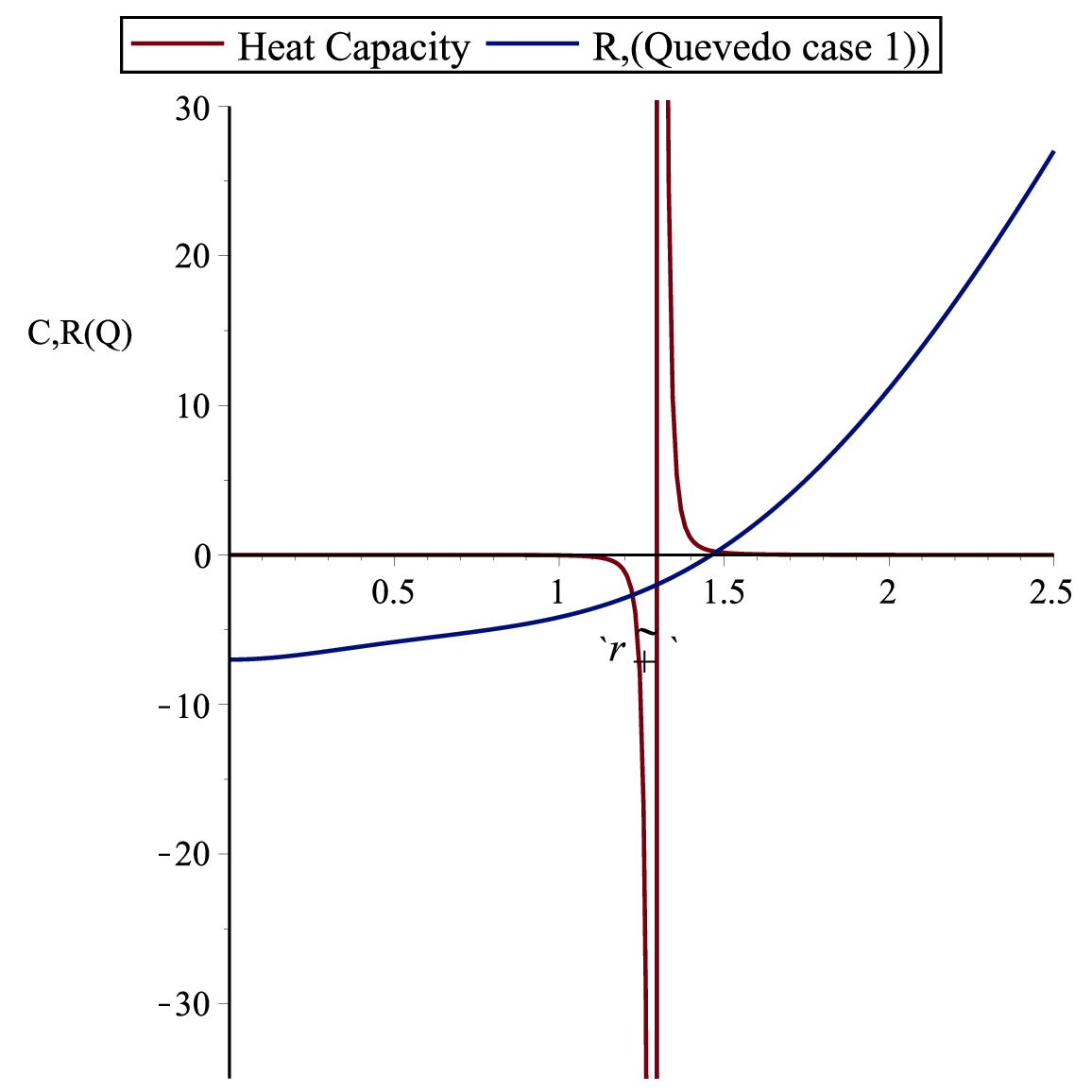}
 \caption{Heat capacity and (Quevedo case 1) at  Q = 4.3,n =0.79 ,a = 0.7 ,y = 97}\label{Fig:8}
\end{minipage}
\hfill
\begin{minipage}[h]{0.43\textwidth}
\includegraphics[width=\textwidth]{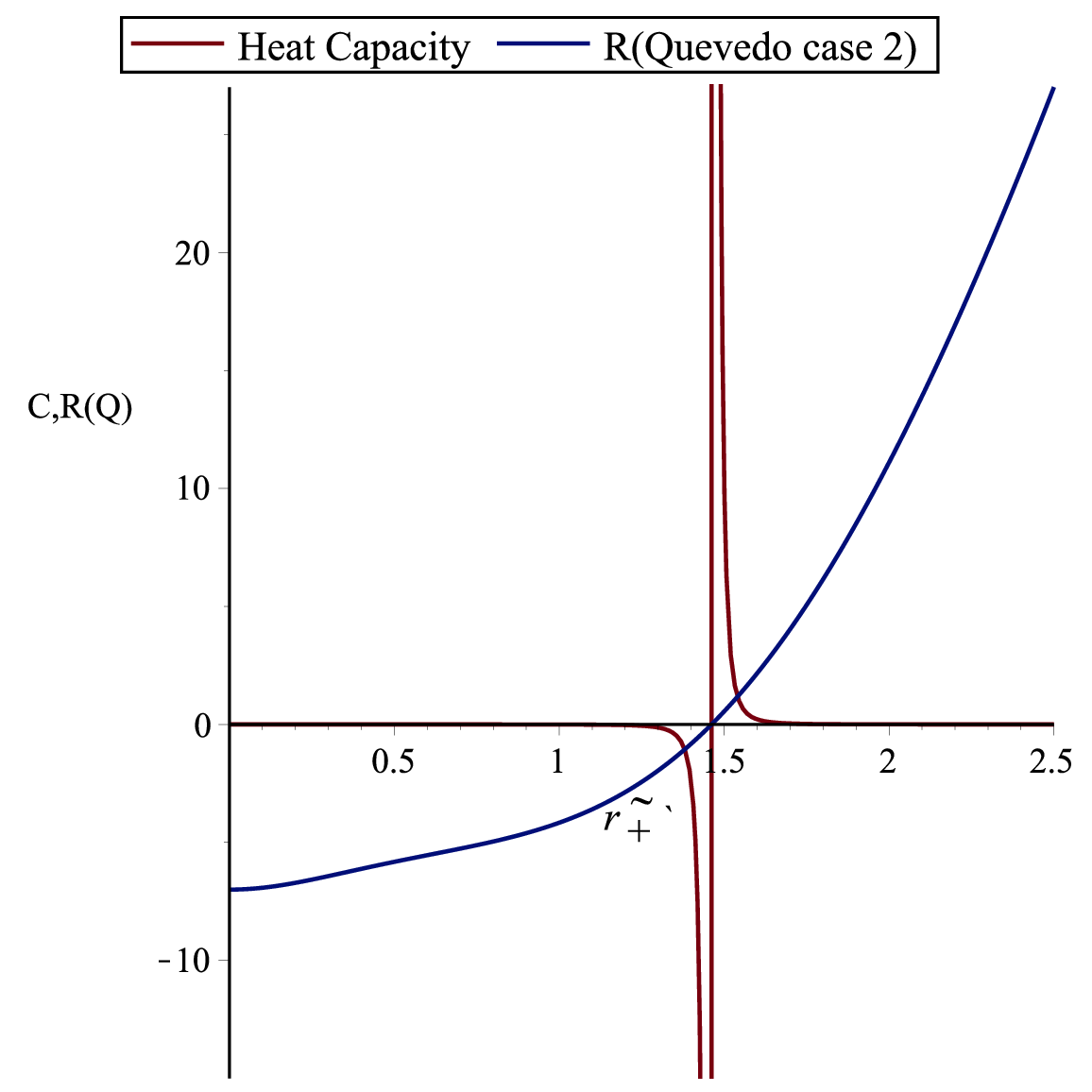}
 \caption{Heat capacity and (Quevedo case 2) at  Q = 4.3,n =0.79 ,a = 0.7 ,y = 97}\label{Fig:9}
\end{minipage}
\end{figure}
\begin{figure}
\centering
\includegraphics[width=0.50\textwidth]{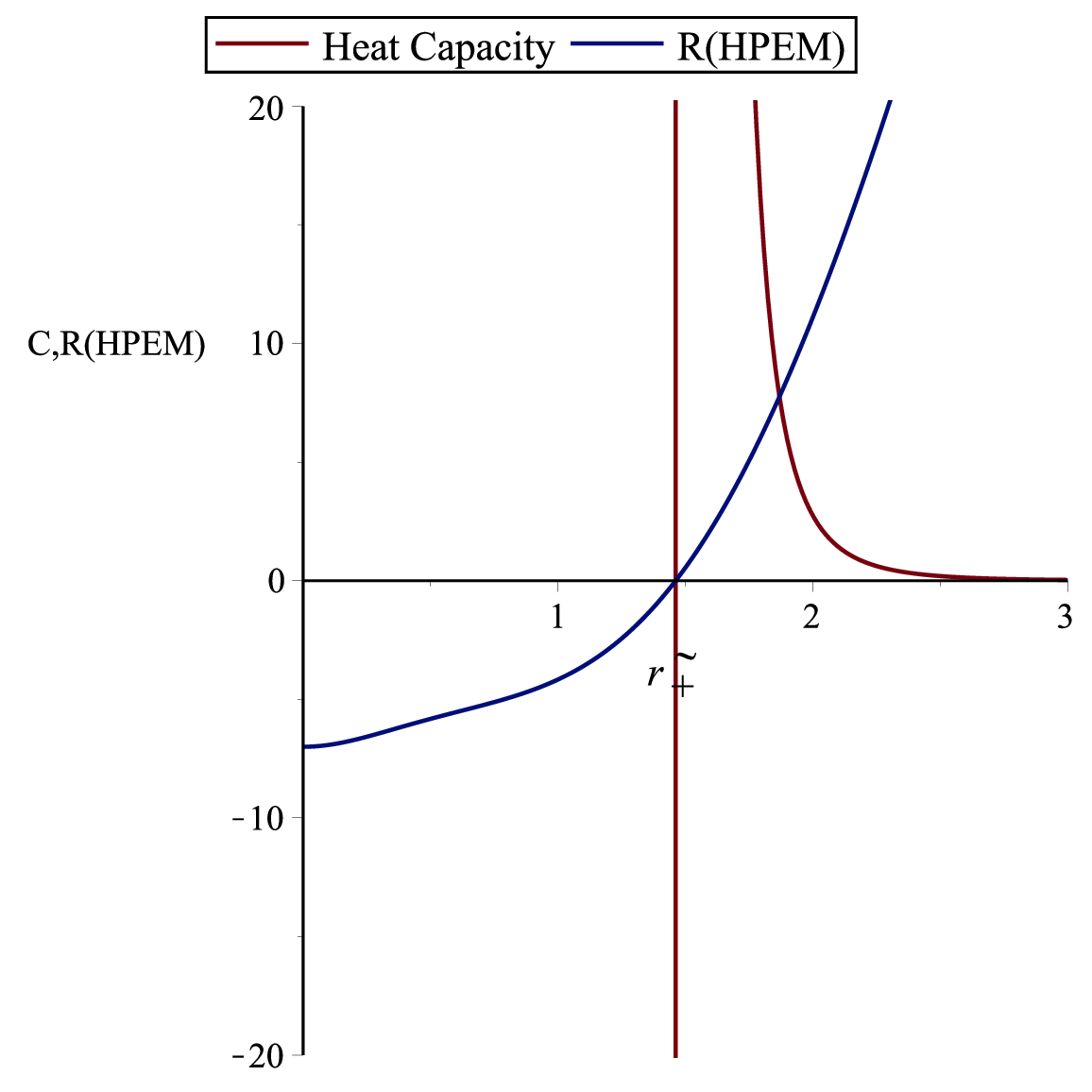}
\caption{Heat capacity and HPEM Metric at Q = 4.3,n =0.79 ,a = 0.7 ,y = 97}\label{Fig:10}
\end{figure}
   The resulting curvature scalars of Ruppeiner, HPEM, and Quevedo (II) are plotted in terms of the   horizon \( \tilde{r}_+ \)  to explore thermodynamic phase transitions, as shown in fig. \ref{Fig:6}, \ref{Fig:9}, \ref{Fig:10}. These graphs show that the curvature scalars for Ruppeiner, HPEM, and Quevedo (II) all have a singular point at \(\tilde{r}_+ = 1.5\). This singular point coincides with the zero point of the heat capacity, indicating a physical limitation. Furthermore, in fig. \ref{Fig:7}, \ref{Fig:8}, we can see that the singular point of the Weinhold and Quevedo (I) curvature scalar does not overlap with the heat capacity divergence points for fixed values  Q = 4.3, n = 0.79, y = 97, and a = 0.7.

   
\section{Conclusion}
In this work, we examined  the shadows and thermal geometries of rotating  charged AdS BH, which has an additional NUT  parameter `n'. Specifically, the effects of spin parameter `a' and NUT parameter`n' on AdS BH shadows are discussed. We find parameters \(\zeta\) and \(\eta\) by applying the circular orbit constraints and construct the null geodesics around the spinning BH [see eqs. (\ref{9}), (\ref{10})]. We have obtained analytical expressions for the shadows cast by an observer at infinity and for particular limits. One can consider that the observer is situated between the cosmological horizon and the outer horizon then by adding the appropriate celestial coordinates to find the BH shadow, which is presented in Figs. \ref{fig:1} and \ref{fig:4}, graphically. One can notice that the BH shadow's size for the observer at infinity increases as NUT charge increases while distorted to D-shaped as \(\theta\) and spin parameter both increase (see figure \ref{fig:1}). For an observer at a particular limit, the shadow's size of BH is larger for higher values of NUT charge, distorted to D-shaped as spin parameter and cosmological constant increase (see figure \ref{fig:4}). \\
Finally, we explore the thermodynamic geometry and calculate the curvature scalar for the Ruppeiner, HPEM, Weinhold, and Quevedo metrics of the corresponding rotating BH. Our findings show that the curvature scalars for the Ruppeiner , HPEM and Quevedo (II) metrics have singular points that align with the zeros of the heat capacity which provide the additional physical information about the interactions in the underlying microscopic statistical basis which are encoded in the scalar curvature of thermodynamic quantities arising from these geometries. Therefore, the scalar curvature is related to the divergency at the critical point and the correlation volume of the system. On the other hand, the singularity of Weinhold and Quevedo (I) metrics do not coincide with zeros of heat capacity which do not provide any physical information. 
\section*{Acknowledgement}
The present research work was funded by Princess Nourah bint Abdulrahman University Researchers Supporting Project number (PNURSP2024R59), Princess Nourah bint Abdulrahman University, Riyadh, Saudi Arabia.
\bibliographystyle{unsrt}

\end{document}